\documentclass[showpacs,superscriptaddress,preprintnumbers,amsmath,amssymb,onecolumn,nofootinbib]{revtex4}

\usepackage{graphicx}
\usepackage{dcolumn}
\usepackage{bm}
\usepackage{type1cm}

\begin{document}

\preprint{KUNS-2141}

\title{Primordial statistical anisotropy generated at the end of inflation}

\author{Shuichiro Yokoyama}
\email{
shu@a.phys.nagoya-u.ac.jp
}
\affiliation{
Department of Physics and Astrophysics, Nagoya University, Aichi 464-8602, Japan
}

\author{Jiro Soda}%
\email{jiro@tap.scphys.kyoto-u.ac.jp}
\affiliation{Department of Physics, Kyoto University, Kyoto 606-8501, Japan}

\date{\today}

\begin{abstract}

We present a new mechanism for generating primordial statistical anisotropy of curvature perturbations.
We introduce a vector field which has a non-minimal kinetic term
and couples with a waterfall field in hybrid inflation model. 
In such a system, the vector field gives fluctuations of the end of inflation
and hence induces a subcomponent of curvature perturbations. 
Since the vector has a preferred direction, the statistical anisotropy
could appear in the fluctuations. We present the explicit formula for 
the statistical anisotropy  in the primordial power
 spectrum and the bispectrum of curvature perturbations.
Interestingly, there is the possibility that
 the statistical anisotropy does not appear in the power spectrum
but does appear in the bispectrum. We also find that the statistical anisotropy
provides the shape dependence to the bispectrum. 

\end{abstract}

\maketitle

\section{Introduction}

Current cosmological observations such as cosmic microwave 
background (CMB) observations tell us that
our universe is likely to be homogeneous and isotropic.
However, there is the possibility that the effect of rotational invariance violation
has been already detected in the CMB temperature anisotropies, as an apparent alignment of the CMB multipoles on very large scales, so-called "Axis of Evil"~\cite{de OliveiraCosta:2003pu,Land:2005ad,Land:2006bn,Hajian:2003qq,Hajian:2006ud,Ackerman:2007nb,ArmendarizPicon:2007nr,ArmendarizPicon:2005jh,Pullen:2007tu,Copi:2005ff,Copi:2006tu,Gumrukcuoglu:2007bx,Gordon:2005ai,Rodrigues:2007ny,Pereira:2007yy,Pitrou:2008gk,Schild:2008fs,Eriksen:2003db,Hansen:2004vq,Jaffe:2005pw,Eriksen:2007pc,Koivisto:2008ig}.
If it is true, the statistical anisotropy might stem from the physics
 in the early universe. Hence, the possibility of statistical anisotropy deserves
 further investigation.

On the theoretical side,  statistically homogeneous, isotropic, and Gaussian
density fluctuations had been supposed to be robust prediction of the inflationary scenario.
However, the possibility of the statistical non-Gaussianity has been 
studied extensively. It turns out that it is not difficult to make a non-Gaussianity if we 
incorporate the multi scalar fields into the system.
 In fact, there are many scalar fields in any realistic particle physics model. 
 Hence, it is natural to test the non-Gaussianity
seriously by making use of observational data.
 For the same reason, in this paper,
we would like to argue that the statistical anisotropy
should be checked observationally. The point is that there are many vector fields
in addition to the scalar fields in the particle physics model. As we will see, 
the vector field could violate the rotational invariance
 at the end of inflation if it couples with a waterfall field in a hybrid inflation model. 
That means the statistical anisotropy could be generated due to the
ubiquitous vector fields. Of course, the curvature perturbations induced by the vector field
should be subdominant in the total curvature perturbations in order to reconcile the scenario
with the current data. Nevertheless, the effect of the vector field
should not be overlooked under the current precision cosmology. 

The paper is organized as follows.
In section~\ref{model}, we show a new mechanism generating the statistical anisotropy
 at the end of inflation.
This model is based on Ref~\cite{Lyth:2005qk,Salem:2005nd,Alabidi:2006wa,Lyth:2006nx,Bernardeau:2002jf,Sasaki:2008uc,Bernardeau:2004zz}, in which the authors consider the
modified hybrid inflation adding a light scalar field other than inflaton 
in order to generate the large non-Gaussianity of primordial curvature perturbations.
We put a massless vector field instead of a light scalar field.
The vector field violates the rotational invariance and as a consequence
generates the statistical anisotropy in the curvature perturbations. 
In section~\ref{spectrum}, we present the calculation of vector fluctuations.
It allows us to explicitly show the effects of violation of rotational invariance 
on the power spectrum and the bispectrum of primordial curvature perturbations.
We also provide a simple example as an illustration. 
We find that the statistical anisotropy in the bispectrum could be more
significant than that in the power spectrum. 
In section~\ref{summary}, we give a brief summary.

\section{Basic idea}
\label{model}

In this section, we give a basic idea of the new mechanism for
 generating the statistical anisotropy in primordial curvature perturbations.
The explicit calculation will be given in the next section. 

In the original hybrid inflation model, 
when the effective mass squared of a waterfall field becomes negative, i.e., the waterfall field becomes tachyonic,
the tachyonic instability makes the waterfall field roll down toward the true vacuum state and
the inflation suddenly ends. The critical value  
$\phi_e$, which is the value of the inflaton $\phi$ at the end of inflation,
should be determined by the model parameter and hence $\phi_e$ does not fluctuate.
However, if some other fields couple with the waterfall field, the situation 
would be different.

Let us consider a system constructed by two scalar fields, say, 
an inflaton $\phi$ and a waterfall field $\chi$, and a vector field 
$A_{\mu} (\mu=0,1,2,3)$ which couples with the waterfall field.
The action can be written as
\begin{eqnarray}
S &\! = \!& {1 \over 2}\int d^4x \sqrt{-g} R 
- \int d^4 x \sqrt{-g} \left[ {1 \over 2}g^{\mu\nu}
\left(\partial_\mu \phi \partial_\nu \phi + \partial_\mu \chi \partial_\nu \chi\right) + V(\phi,\chi,A_\mu)\right] \nonumber\\
&&
\qquad\qquad\qquad
-{1 \over 4}\int d^4 x \sqrt{-g}g^{\mu\nu}g^{\rho\sigma}f^2(\phi) F_{\mu\rho}F_{\nu\sigma}~,
\end{eqnarray}
where $F_{\mu\nu} \equiv \partial_\mu A_\nu - \partial_\nu A_\mu$ is the field strength
of the vector field, $V(\phi,\chi,A_\mu)$ is the potential of fields,
and an arbitrary function $f(\phi)$ represents gauge coupling.
 We have to choose $V(\phi,\chi,A_\mu)$ so as not to violate the gauge invariance.
  For the background universe to be isotropic,
we need the condition that
the energy of the inflaton should be dominant in the total energy of the universe
during inflation era. Under this condition, we have a small expectation value of the vector field. 
Here, we do not specify the form of potential but we assume that the vector field is massless
 and have a small expectation value compared with the inflaton.
Hence, we neglect the terms which come from the coupling with the vector field
in the background equations of motion for the scalar fields
and
we treat the vector field perturbatively.

\subsection{$\delta N$ formalism}
\label{deltaN}

First, we briefly review the $\delta N$ formalism~\cite{Sasaki:1995aw,Sasaki:1998ug,Lyth:2004gb,Lyth:2005fi}
(or separate universe approach~\cite{Wands:2000dp}), which is well-known as 
 the most powerful tool to calculate the evolution of curvature perturbations on super-horizon scales.
Based on the $\delta N$ formalism, the curvature perturbations on super-horizon scales, where spatial gradients and anisotropic shear are small, 
can be written by 
\begin{eqnarray}
\zeta (t_F) \simeq \delta N (t_F,t_*)~,
\label{zetadelta}
\end{eqnarray}
where $\delta N (t_F,t_*)$ represents the perturbation of the $e$-folding number 
 $N$ measured in homogeneous background universe between the initial flat hypersurface
  at $t=t_*$ and the final uniform energy density hypersurface at $t=t_F$. 
More precisely, the $e$-folding number is given by $N=\int^{t_F}_{t_*} H dt~,$ 
 with the Hubble parameter $H$. 
When we take the final time $t_F$ to be a certain time after the
complete convergence of the background trajectories in phase space,
$\zeta$ becomes constant.
Let the initial time $t_*$ be some time during the slow-roll inflation driven by the inflaton.
 Soon after the time that the scale of interest crosses the horizon,
  we can rewrite Eq.~(\ref{zetadelta}) as a Taylor expansion up to the second order
\begin{eqnarray}
\zeta (t_F) = N_{I*} \delta\phi^I_* + {1 \over 2}N_{IJ*}\delta \phi^I_* \delta \phi^J_*~,
\end{eqnarray}
where $N_{I*} \equiv \partial N(t_F,t_*)/\partial \phi^I(t_*)$ and
$N_{IJ*} \equiv \partial^2 N(t_F,t_*)/\partial \phi^I(t_*)\partial\phi^J(t_*)$.
Here, we consider that the background trajectories are parameterized by the scalar field $\phi^I$.
Hence, the perturbations of the scalar field $\delta \phi^I$ give rise to
 the difference between the evolutions of
the adjacent background homogeneous trajectories.

\subsection{Curvature perturbation generated at the end of inflation}

As we mentioned in the previous subsection,
the curvature perturbations on super-horizon scales are
 given by the perturbations of e-folding number $\delta N$.
In the standard single scalar inflation or hybrid inflation, the
inflation ends when the inflaton $\phi$ reaches a critical value $\phi_e$.
On the other hand, in the multi-component inflation, the critical value
$\phi_e$ may fluctuate due to a light field other than the inflaton $\phi$.
Hence, in such a situation, there is a possibility of generating the curvature perturbations
through the fluctuation of $\phi_e$.
Based on the models introduced by
Ref.~\cite{Lyth:2005qk,Alabidi:2006wa},
in this paper, we introduce a massless vector field $A^\mu$ as another light field.
Using the $\delta N$ formula,
the curvature perturbation generated at the end of inflation ($t=t_{e}$) can be expressed as 
\begin{eqnarray}
\zeta_{end} &=& {\partial N \over \partial \phi_e} \delta \phi_e
+ {1 \over 2}{\partial^2 N \over \partial \phi^2_e} \delta \phi_e^2 \nonumber\\
&=& {\partial N \over \partial \phi_e} {d \phi_e(A) \over d A^i}\delta A^i + {1 \over 2}
\left[{\partial N \over \partial \phi_e} {d^2 \phi_e(A) \over d A^i d A^j} + {\partial^2 N \over \partial \phi^2_e} {d \phi_e(A) \over d A^i}{d \phi_e(A) \over d A^j}\right]\delta A^i \delta A^j~,
\end{eqnarray}
where we set $A^0=0$.
That is, $\phi_e$ can fluctuate due to fluctuations of the vector field $A^i$.
Let us take the hypersurface at the end of inflation $t=t_e$ to be that of 
a uniform energy density. Then, 
the total curvature perturbations at the end of inflation $t=t_{e}$ is given by
\begin{eqnarray}
\zeta (t_{e}) = \zeta_{inf} + \zeta_{end}~,
\end{eqnarray}
where
\begin{eqnarray}
&&\zeta_{inf} = {\partial N \over \partial \phi_*}\delta \phi_* + {1 \over 2}{\partial^2 N \over \partial \phi^2_*}\delta \phi_*^2~.
\end{eqnarray}
We assume that all fields decay into radiation component
 through the sudden reheating just after the end of inflation.
Under this assumption, the evolution of the universe can be described by one-parameter, namely, radiation energy density, after the end of inflation. 
Hence, the curvature perturbations on a uniform energy density hypersurface
become constant after the end of inflation and the iso-curvature mode does not persist until present.
At the leading order, the power spectrum of curvature perturbations is given by
\begin{eqnarray}
\langle \zeta_{{\bf k}_1} \zeta_{{\bf k}_2}\rangle =
P_\zeta({\bf k}_1)\delta^{(3)}\left({\bf k}_1 + {\bf k}_2\right)= N_*^2 P_\phi(k_1)
\delta^{(3)}\left({\bf k}_1 + {\bf k}_2\right)
+ N_e^2 {d \phi_e(A) \over d A^i}{d \phi_e(A) \over d A^j}
 \langle \delta A^i_{{\bf k}_1} \delta A^j_{{\bf k}_2}\rangle~,
 \label{curvps}
\end{eqnarray}
where $N_* = \partial N /\partial \phi_*$, $N_e = \partial N / \partial \phi_e$
and $\langle \delta A^i_{{\bf k}_1} \delta A^j_{{\bf k}_2}\rangle$ is the power spectrum of the perturbation of the vector field, which we will calculate in the next section.
Here, we neglect the cross terms such as 
$\langle \delta \phi_{{\bf k}_1}  \delta A^i_{{\bf k}_2}\rangle$ and 
assume that $\delta \phi_*$ has pure Gaussian distributions.
To the leading order, the bispectrum is also given by
\begin{eqnarray}
\langle \zeta_{{\bf k}_1} \zeta_{{\bf k}_2} \zeta_{{\bf k}_3}\rangle
&\! \equiv &\! (2\pi)^{-3/2}B_\zeta({\bf k}_1, {\bf k}_2, {\bf k}_3) \delta^{(3)}\left({\bf k}_1 + {\bf k}_2 + {\bf k}_3 \right) \nonumber\\
&\!=&\! (2\pi)^{-3/2}N_*^2 N_{**} \left[P_\phi(k_1)P_\phi(k_2) + 2~ {\rm perms} \right] \delta^{(3)}\left({\bf k}_1 + {\bf k}_2 + {\bf k}_3 \right)\nonumber\\
&& 
+ N_e^3 {d \phi_e(A) \over d A^i}{d \phi_e(A) \over d A^j}{d \phi_e(A) \over d A^\ell}
 \langle \delta A^i_{{\bf k}_1} \delta A^j_{{\bf k}_2} \delta A^\ell_{{\bf k}_3}\rangle \nonumber\\
&& 
+  N_e^4 {d \phi_e(A) \over d A^i}{d \phi_e(A) \over d A^j}
\left({1 \over N_e}{d^2 \phi_e(A) \over d A^{\ell_1}dA^{\ell_2}}+{N_{ee} \over N_e^2}
 {d \phi_e(A) \over d A^{\ell_1}}{d \phi_e(A) \over d A^{\ell_2}}
\right) \nonumber\\
&&\qquad\qquad\qquad\qquad\qquad\qquad \times
\left[\langle  \delta A^i_{{\bf k}_1} \delta A^j_{{\bf k}_2}
\left(\delta A^{\ell_1} \star \delta A^{\ell_2}\right)_{{\bf k}_3} \rangle  + 2~ {\rm perms}\right]~, \nonumber\\
\label{curvbi}
\end{eqnarray}
where $N_{**} = \partial^2 N /\partial \phi_*\partial \phi_*$, $N_e = \partial^2 N / \partial \phi_e \partial \phi_e$ and $\star$ denotes the convolution.

From these expressions, we find that one can obtain the curvature perturbations 
which have the direction-dependence due to
the vector field perturbations.

\section{Primordial Statistical Anisotropy}
\label{spectrum}

In this section, we first quantize the vector field with a non-minimal kinetic term
and obtain the power spectrum of the vector fluctuations.
 Next, we derive the explicit formula for the power
spectrum and the bispectrum of curvature perturbations. 
Then, we present a simple example as an illustration. 

\subsection{The spectrum of vector perturbations}
\label{vecpert}
Here, we give the spectrum of vector perturbations, following Ref.~\cite{Martin:2007ue}, which appear in the expression of
the spectrum of curvature perturbations~(\ref{curvps}) and (\ref{curvbi}).

Following Ref.~\cite{Martin:2007ue},
we adopt the Coulomb gauge where $\delta A_0(x^\mu)=0$ and $\partial_jA^j(x^\mu)=0$.
Then, we obtain the evolution equation for the linear perturbation as
\begin{eqnarray}
{\cal A}''_i - {f'' \over f}{\cal A}_i - a^2\partial_j\partial^j{\cal A}_i = 0~,
\end{eqnarray}
where ${\cal A}_i \equiv f \delta A_i$, a prime represents the derivative with respect to a conformal time, $\tau=\int dt/a$,
and we neglected the potential term.
Here, we assumed the gauge invariance for the vector field. Hence, there should be
a derivative coupling between the vector field and the waterfall field.
However, this term is not relevant to our analysis.

Now, we can expand the vector field in terms of annihilation and creation operators, $\hat{a}_{\lambda{\bf k}}$ and $\hat{a}^{\dagger}_{\lambda{\bf k}}$, as 
\begin{eqnarray}
{\cal A}_i(\tau, x^i) = \int {d^3 k \over \left(2\pi\right)^{3/2}}\sum_{\lambda=1}^{2}
\epsilon_{i\lambda}({\bf k})\left[v_k(\tau)\hat{a}_{\lambda{\bf k}}e^{i{\bf k}\cdot{\bf x}}
+v^*_k(\tau)\hat{a}^{\dagger}_{\lambda{\bf k}}e^{-i{\bf k}\cdot{\bf x}}\right]~,
\end{eqnarray}
where ${\bf k}$ is the comoving wave number
and the transverse polarization vectors $\epsilon_{i \lambda}$ defined by
\begin{eqnarray}
\sum_{\lambda=1}^{2}\epsilon^{i}_{\lambda}({\bf k})\epsilon_{j\lambda}({\bf k}) 
= \delta^i_{~j} - \delta_{j\ell}{k^ik^\ell \over k^2} ~,
\end{eqnarray}
 are introduced for a consistent quantization in the Coulomb gauge.
With the change of the variable $\psi_k = a v_k$,
the equation of motion for the mode function $\psi_k$ is given by
\begin{eqnarray}
\psi''_k + \left( k^2 - {f'' \over f}\right)\psi_k = 0 \ .
\label{quanteom}
\end{eqnarray}
Using this variable, we obtain the power spectrum of vector perturbations as
\begin{eqnarray}
\langle \delta A_i({\bf k}) \delta A_j({\bf k}') \rangle = {\left|\psi_k\right|^2 \over a^2f^2}\left(\delta_{ij} - {k_ik_j \over k^2}\right)\delta^{(3)}({\bf k} + {\bf k}')~,
\label{ps}
\end{eqnarray}
where $k_i = \delta_{ij}k^j$.
In the short wavelength limit $k^2 \to \infty$,
from Eq.~(\ref{quanteom}), we see one can take the Bunch-Davies vacuum
\begin{eqnarray}
\psi_k \to {1 \over \sqrt{2k}}e^{-ik\tau}~.
\label{BD}
\end{eqnarray}
In the slow-roll limit $(\epsilon = -\dot{H}/H^2 \to 0)$, we have
\begin{eqnarray}
- \tau ={1 \over aH}\left(1 + O(\epsilon)\right)~.
\end{eqnarray}
Hence, we assume that $d\tau = da/(a^2 H)$ and
assuming that the form of the gauge coupling function $f \propto a^\alpha$,
we have
\begin{eqnarray}
\psi''_k + \left( k^2 - {\alpha (\alpha + 1) \over \tau^2} \right)\psi_k = 0~.
\end{eqnarray}
With the Bunch-Davies vacuum (\ref{BD}), this equation can be solved easily as
\begin{eqnarray}
\psi_k = \left({\pi \over 4k}\right)^{1 \over 2}\exp \left[i(\alpha+1) {\pi \over 2} \right](-k\tau)^{1/2}
H^{(1)}_{\alpha + 1/2}(-k\tau)~,
\end{eqnarray}
where $H^{(1)}_{\nu}(x)$ is the Hankel function of the first kind.
If we take $\alpha = 1$ or $-2$, 
in the long wavelength limit $-k\tau \to 0$, we obtain the asymptotic solution for $\psi_k$ as
\begin{eqnarray}
\psi_k (\alpha = 1,-2) \to {\left(-ik\tau\right)^{-1} \over \sqrt{2k}}\left[1+O\left(\left(-k\tau\right)^2\right)
\right]~.
\end{eqnarray}
Thus, we can obtain the scale-invariant vector power spectrum
\begin{eqnarray}
\langle \delta A^i_{{\bf k}} \delta A^j_{{\bf k}'} \rangle &\!=\!& {H^2_{\rm inf} \over 2k^3f^2}\left(\delta^{ij} - {k^ik^j \over k^2}\right)\delta^{(3)}({\bf k} + {\bf k}') \nonumber\\
&\!=\!& P_\phi(k)f^{-2}
\left(\delta^{ij} - {k^ik^j \over k^2}\right)\delta^{(3)}({\bf k} + {\bf k}')~,
\label{finalpsvec}
\end{eqnarray}
 where we defined  $P_\phi(k) = {H_{\rm inf}^2}/2k^3$ and $H_{\rm inf}$ denotes the Hubble parameter
 during the inflation.
Strictly speaking, in order to evaluate the bispectrum we need to calculate
$\langle \delta A^i_{{\bf k}_1} \delta A^j_{{\bf k}_2} \delta A^\ell_{{\bf k}_3}\rangle$.
However, as we saw in this subsection, 
initially the vector perturbation in Coulomb gauge can be considered as
the Gaussian perturbation as well as the perturbation of scalar field in the standard slow-roll inflation model. 
In the standard slow-roll inflation model, as is well known,
three point correlation function of the perturbation of scalar field 
is suppressed by the slow-roll parameters~\cite{Maldacena:2002vr}, that is, we can consider that
the perturbations of scalar field at the horizon crossing time $t=t_*$ almost obey Gaussian statistics.  
In this model, the vector field can be considered to be slowly rolling, so
the deviation from Gaussian statistics of the vector perturbations $\delta A^i$
would be too small to be detected 
in the future experiments, e.x., Planck~\cite{Komatsu:2001rj,:2006uk}.
Hence, 
we assume, here, that the non-Gaussian feature of the vector field perturbation is negligible.
 
\subsection{Rotation Violating Power spectrum and Bispectrum}

Now, we are in a position to write down the formula for the power spectrum
and the bispectrum of curvature perturbations. 

Using Eq.~(\ref{finalpsvec}), 
we can rewrite the expression for the power spectrum of curvature perturbations
in the modified hybrid inflation model (\ref{curvps}) as
\begin{eqnarray}
\langle \zeta_{{\bf k}_1} \zeta_{{\bf k}_2}\rangle =
P_\zeta({\bf k}_1)\delta^{(3)}\left({\bf k}_1 + {\bf k}_2\right)= 
\left[ N_*^2+ \left({N_e \over f_e}\right)^2 {q_i}{q_j} 
\left(\delta^{ij} - {k^i_1k^j_1 \over k^2_1}\right)
\right]
 P_\phi(k_1)
\delta^{(3)}\left({\bf k}_1 + {\bf k}_2\right)~,
\end{eqnarray}
where we have defined $q_i \equiv d\phi_e/dA^i$, $q_{ij} \equiv d^2\phi_e/dA^idA^j$.
In order to compare this expression with that of Ref.~\cite{Ackerman:2007nb},
$P_{\zeta}$ can be rewritten as
\begin{eqnarray}
 P_\zeta({\bf k}) = P^{iso}_{\zeta}(k)\left[ 1 + g_\beta (\hat{\bf q}\cdot \hat{\bf k})^2 \right]~,
\end{eqnarray}
where we used the notation $\hat{\bf q} = {\bf q}/\left|{\bf q}\right|$, $\hat{\bf k} = {\bf k}/\left|{\bf k}\right|$ and the isotropic part
\begin{eqnarray}
P^{iso}_{\zeta}(k) = {P_\phi(k) \over 2\epsilon_*}(1+\beta)~,
\label{anisp}
\end{eqnarray}
is separated. The coefficient of the anisotropic part reads 
\begin{eqnarray}
~g_\beta = - {\beta \over 1+\beta}~, \quad 
\beta \equiv \left({N_e \over N_* f_e}\right)^2\left|{\bf q}\right|^2~,
\end{eqnarray}
where we defined $N_* = 1/\sqrt{2\epsilon_*}$.
Here, $\epsilon \equiv V_\phi^2/2V^2$
 is a slow-roll parameter.
 
The bispectrum  (\ref{curvbi}) also becomes
\begin{eqnarray}
&&B_\zeta({\bf k}_1, {\bf k}_2, {\bf k}_3)
= \Biggl\{\Biggl[
N_*^2N_{**}+ \left({N_e \over f_e}\right)^4{q_i}{q_j}
\left({1 \over N_e}{q_{\ell_1\ell_2}}
+ {N_{ee} \over N_e^2}
{q_{\ell_1}}{q_{\ell_2}}
\right)
\left(\delta^{i\ell_1} - {k^i_1k^{\ell_1}_1 \over k^2_1}\right)
\left(\delta^{j\ell_2} - {k^j_2k^{\ell_2}_2 \over k^2_2}\right)
 \Biggr] \nonumber\\
&& \qquad\qquad\qquad\qquad\qquad\qquad\qquad\qquad\qquad\qquad\qquad\qquad\qquad\qquad
\times P_\phi(k_1)P_\phi(k_2) 
+ 2~{\rm perms}\Biggr\}~,\nonumber\\
\end{eqnarray}
where  we assumed that $\delta A_i$ is  Gaussian.
Now, we define the non-linear parameter $f_{NL}$ as the bispectrum normalized by the isotropic part of power spectrum
$P_\zeta^{iso}(k)$; 
\begin{eqnarray}
{6 \over 5}f_{NL}({\bf k}_1,{\bf k}_2,{\bf k}_3)
&\!\equiv \!& {B_\zeta({\bf k}_1,{\bf k}_2,{\bf k}_3) \over P_\zeta^{iso}(k_1)P_\zeta^{iso}(k_2) + 2~{\rm perms}}
=
\left[
{F({\bf k}_1,{\bf k}_2)k^3_3 \over \sum_{i}{ k_i^3}}
+ 2~{\rm perms}\right]~, \nonumber\\
\end{eqnarray}
where we assumed the scale-invariant power spectrum and defined
\begin{eqnarray}
F({\bf k}_1,{\bf k}_2) &\!\equiv\!& (1+\beta)^{-2}{N_{**} \over N_*^2} + g_\beta^2{N_{ee} \over N_e^2}
\left[ 1 - (\hat{\bf q} \cdot \hat{\bf k}_1)^2 \right]\left[ 1 - (\hat{\bf q} \cdot \hat{\bf k}_2)^2 \right] \nonumber\\
&& \qquad\qquad
+{g_\beta^2 \over N_e}{q_{\ell_1\ell_2} \over  \left|{\bf q}\right|^2}\left[ \hat{q}^{\ell_1} - (\hat{\bf q} \cdot \hat{\bf k}_1)\hat{k}^{\ell_1}_1 \right]\left[\hat{q}^{\ell_2} - (\hat{\bf q} \cdot \hat{\bf k}_2)\hat{k}^{\ell_2}_2 \right]~.
\label{26}
\end{eqnarray}
Strictly speaking, we should have defined the non-linear parameter
by using the full power spectrum instead of the isotropic one.
However, the resultant expression is too complicated to display here.
When the anisotropic part in the power spectrum is small, 
the above definition gives a good approximation
to the true non-linear parameter. Indeed, we will discuss this interesting possibility later.  
Otherwise, we should go back to the original formula for the bispectrum.

Under the slow-roll approximation, we obtain $N_{**}/N_*^2 \sim N_{ee}/N_e^2 = O(\epsilon)$.
Hence, neglecting the first and second terms in the right hand side of the above equation,
we can further simplify the expression (\ref{26}) as
\begin{eqnarray}
F({\bf k}_1, {\bf k}_2) &\simeq& {g_\beta^2 \over N_e}{\hat{q}_{\ell_1\ell_2}}\left[ \hat{q}^{\ell_1} - (\hat{\bf q} \cdot \hat{\bf k}_1)\hat{k}^{\ell_1}_1 \right]\left[\hat{q}^{\ell_2} - (\hat{\bf q} \cdot \hat{\bf k}_2)\hat{k}^{\ell_2}_2 \right]~,
\end{eqnarray}
where $\hat{q}_{\ell_1 \ell_2} = {q}_{\ell_1 \ell_2}/\left|{\bf q}\right|^2$.
We can decompose the non-linear parameter into the isotropic part and the anisotropic part as
\begin{eqnarray}
{6 \over 5}f_{NL}({\bf k}_1, {\bf k}_2, {\bf k}_3)
&\! = \!&
{6 \over 5}f_{NL}^{iso}\left( 1 + f^{ani}( {\bf k}_1, {\bf k}_2, {\bf k}_3) \right)~,
\label{anisofnl}
\end{eqnarray}
where the isotropic part reads
\begin{eqnarray}
{6 \over 5}f_{NL}^{iso} \equiv {\sqrt{2 \epsilon_e} g_\beta^2}
\hat{q}_{ij}\hat{q}^i\hat{q}^j~, \label{fnliso}
\end{eqnarray}
and the anisotropic part is deduced as
\begin{eqnarray}
f^{ani}({\bf k}_1,{\bf k}_2, {\bf k}_3) = {\hat{q}_{ij} \over \hat{q}_{kl}\hat{q}^k\hat{q}^l} 
\left\{ \left[
- \hat{q}^i\left(\hat{k}_1^j (\hat{\bf q}\cdot \hat{\bf k}_1) 
+ \hat{k}_2^j (\hat{\bf q}\cdot \hat{\bf k}_2)
\right)
+
\hat{k}^i_1\hat{k}^j_2 (\hat{\bf q}\cdot \hat{\bf k}_1) 
(\hat{\bf q}\cdot \hat{\bf k}_2) \right]
{k_3^3
 \over \sum_{i}{ k_i^3}}
+ 2~{\rm perms}\right\}~.\label{fnlani}
\end{eqnarray}
Here, we used the relation $N_e = 1/\sqrt{2\epsilon_e}$.
Taking a look at the above formula, we notice that the statistical anisotropy
gives the shape to the bispectrum even for the local model.

\subsection{Simple example}

Let us consider a simple example given by
\begin{eqnarray}
V(\phi,\chi,A^i) = {\lambda \over 4}\left(\chi^2 - v^2\right)^2 + {1 \over 2}g^2 \phi^2 \chi^2 + {1 \over 2}m^2\phi^2
+ {1 \over 2}h^2 A^\mu A_\mu \chi^2~,
\end{eqnarray}
where $\phi$, $\chi$ and $A^\mu$ are  the inflaton, the waterfall field,
 and the light vector field, respectively.
The coupling constants are denoted by $\lambda, g, h$, the inflaton mass is given by $m$,
and the vacuum expectation value for $\chi$ is represented by $v$.
The effective mass term of the vector can be accommodated in the system without
violating the gauge invariance. 
Indeed, the potential (31) looks like an abelian Higgs model in unitary gauge. 
Namely, $\chi$ could be the radial direction of a complex scalar field
which is invariant under a local $U(1)$ gauge transformation and $\phi$ 
would be a gauge singlet.
For this potential, we can realize the hybrid inflation
and obtain the evolution of inflaton field, $\phi$, as
\begin{eqnarray}
\phi \propto a^{-m^2/V_0}~,
\end{eqnarray}
where $V_0 = \lambda v^4/4$.
 Hence, when $f(\phi) \propto \phi^{2V_0/m^2}$, we have $f \propto a^{-2}$.
This behavior was assumed in the previous subsection \ref{vecpert} 
in order to achieve the scale-invariant vector power spectrum.
The effective mass squared of the waterfall field $\chi$ is given by
\begin{eqnarray}
m^2_\chi \equiv - \lambda v^2 + g^2 \phi^2 + h^2 A^iA_i~,
\end{eqnarray}
where we set $A^{0}=0$.
At the end of inflation, the equality;
\begin{eqnarray}
\lambda v^2 = g^2 \phi^2_e + h^2 A^i A_{i}~,
\end{eqnarray}
is satisfied. As is expected, the critical value, $\phi_e$, depends on $A^i$.
Hence, we have
\begin{eqnarray}
\hat{q}^i = - { A^i \over \left|{\bf A}\right|}~,
~\hat{q}^{ij} = -{1 \over \phi_e}\left({g^2 \phi_e^2 \over h^2 \left|{\bf A}\right|^2} \delta^{ij} + \hat{q}^i \hat{q}^j \right)~,
\label{form1}
\end{eqnarray}
and
\begin{eqnarray}
\beta \simeq {1 \over f_e^2}\left({h^2\left|{\bf A}\right| \over g^2 \phi_e}\right)^2~,
\label{form2}
\end{eqnarray}
where we used the approximation $N_* \simeq N_e$.

Now, we can write down the power spectrum
\begin{eqnarray}
 P_\zeta({\bf k}) = P^{iso}_{\zeta}(k)\left[ 1 
 - \frac{{1 \over f_e^2}\left({h^2\left|{\bf A}\right| \over g^2 \phi_e}\right)^2}
 {1+{1 \over f_e^2}\left({h^2\left|{\bf A}\right| \over g^2 \phi_e}\right)^2} 
 (\hat{\bf q}\cdot \hat{\bf k})^2 \right]~.
\end{eqnarray}
It should be noted that the power spectrum is scale invariant although
there is a direction dependence. 
The magnitude of statistical anisotropy is determined
 by the parameter, $g_\beta$.
 
Let us move on to the bispectrum which is more interesting. 
Substituting the expressions (\ref{form1}) and (\ref{form2}) 
into Eqs.~(\ref{fnliso}) and (\ref{fnlani}), 
we have the isotropic part of the non-linear parameter
\begin{eqnarray}
{6 \over 5}f_{NL}^{iso} \simeq
 -\eta_e g_\beta^2\left(1 + {g^2 \phi_e^2 \over h^2 \left|{\bf A}\right|^2}\right) \ ,
\label{fnliso2}
\end{eqnarray}
where $\eta = V_{\phi\phi}/V$.  The anisotropic part reads
\begin{eqnarray}
f^{ani}({\bf k}_1,{\bf k}_2,{\bf k}_3) &\!
=\!& -\left\{ \left[\left(\hat{\bf q}\cdot \hat{\bf k}_1\right)^2
 + \left(\hat{\bf q}\cdot \hat{\bf k}_2\right)^2\right]{k_3^3
 \over \sum_{i}{ k_i^3}}
 + 2~{\rm perms} \right\}\nonumber\\
&&
+ \left(1 + {g^2 \phi_e^2 \over h^2 \left|{\bf A}\right|^2}\right)^{-1}\left\{
\left[\left(\hat{\bf q}\cdot \hat{\bf k}_1\right)^2\left(\hat{\bf q}\cdot \hat{\bf k}_2\right)^2 + {g^2 \phi_e^2 \over h^2 \left|{\bf A}\right|^2}\left(\hat{\bf k}_1\cdot \hat{\bf k}_2\right)\left(\hat{\bf q}\cdot \hat{\bf k}_1\right)\left(\hat{\bf q}\cdot \hat{\bf k}_2\right) \right]{k_3^3
 \over \sum_{i}{ k_i^3}}
+ 2~{\rm perms}\right\}~,\nonumber\\
\label{fani}
\end{eqnarray}
where the relation between wavenumber vectors holds $\hat{\bf k}_1+\hat{\bf k}_2+\hat{\bf k}_3 = 0$.
From the above expression, we see that 
the amplitude of the non-Gaussianity $f_{NL}^{iso}$ is dependent on 
the magnitude of statistical anisotropy, $g_\beta$.
For large $\beta$ ($\beta \gg 1$), we can approximate $g_\beta \simeq 1$, and then,
the statistical anisotropy appearing in the primordial power spectrum (\ref{anisp}) becomes large.
On the other hand, for small $\beta$ ($\beta \ll 1$), 
we can approximate $g_\beta \simeq \beta \ll 1$.
Naively, in this limit, the statistical anisotropy is small and the non-linear parameter $f_{NL}$   is also too small to be detected in the future experiments.
However,  Eq.~(\ref{fnliso2}) tells us that
 there may be a possibility of generating large
 non-Gaussianity even for the small statistical anisotropy in the power spectrum 
 if we choose a small value for $f_e$ and much large value for 
${g^2 \phi_e^2 \over h^2 \left|{\bf A}\right|^2}$ while keeping $g_\beta \ll 1$.
From Eqs.~(\ref{anisofnl}) and (\ref{fani}),
we also find that the anisotropic part has
the same order as the isotropic part in the bispectrum,
in contrast to the power spectrum.
Hence, it may be possible to detect the statistical anisotropy
in the bispectrum with the future experiments.
It will give us information about
a new physics in the early universe associated with the violation
of the rotational invariance.

In this model, it is important to set the initial value of
the vector field to be small.
In general, the vector field $A^\mu$
 may not be trapped near the ground state.
In such a case, however, the prediction contradicts the observations, that is,
this case predicts the anisotropic universe.
Hence, it is not our universe. Of course, there may be more 
attractive scenario which does not resort to the anthropic argument.

\section{summary}
\label{summary}

We presented a new mechanism for generating primordial statistical 
anisotropy of curvature perturbations.
We have considered a vector field coupled with a waterfall field in hybrid inflation model.
We have shown that, in such a system, the statistical anisotropy could appear 
not only in the primordial power
 spectrum but also in the bispectrum of curvature perturbations.
From the formula for the power spectrum and bispectrum of the curvature perturbations,
we see the magnitude of the direction-dependence is related to that of 
non-Gaussianity of curvature perturbations.
The degree of the anisotropy depends on the coupling parameter 
between the vector field and the waterfall field
and also the coupling between the inflaton field and the waterfall field.
Intriguingly, it turned out that there is the possibility that
 the statistical anisotropy does not appear in the power spectrum
but does appear in the bispectrum.
It is also interesting to observe that the shape of the bispectrum
changes due to the statistical anisotropy. It is known that the shape of the
bispectrum is different from model to model. The statistical
anisotropy may have to be taken into account when the data is interpreted. 

Recently, many authors discussed the statistical anisotropy of CMB temperature anisotropies, so-called "Axis of Evil".
The resulting expressions for the spectrum of curvature perturbation, 
(\ref{anisp}), (\ref{fnliso}) and (\ref{fnlani})
can be compared with observations.
If this statistical anisotropy were detected significantly,
the mechanism presented in this paper would explain such effects without contradicting 
other current observations, just by tuning the coupling parameters.

In this paper, in order to obtain the spectrum of perturbations of vector field 
we considered a system
 in which conformal invariance is violated based on Ref.~\cite{Martin:2007ue}
 and assumed that $\alpha = 1$ which represents the dependence of coupling function $f$ on a scale factor $a$. We think that other forms of function $f$ are also possible. It would be interesting to investigate those cases. 
We can also consider  
other systems where the conformal invariance 
is broken~\cite{Golovnev:2008cf,Koivisto:2008xf,
Dimopoulos:2008rf,Dimopoulos:2006ms,Dimopoulos:2007zb}.
The model we presented in this paper is just a toy model.
We need to investigate realistic models based on 
high energy physics which can accommodate our new mechanism.
This would be an issue for future work.

\begin{acknowledgments}

We are grateful to D. Lyth and K. M. Gorski for useful comments.
S.Y. is supported in part by Grant-in-Aid for Scientific Research
on Priority Areas No. 467 ``Probing the Dark Energy through an
Extremely Wide and Deep Survey with Subaru Telescope'', by the
Mitsubishi Foundation,
and by Japan Society for Promotion of Science (JSPS) Core-to-Core
Program ``International Research Network for Dark Energy'', and by
Grant-in-Aids for Scientific Research
(Nos.~18740132,~18540277,~18654047).  
J.S. is supported by the Japan-U.K. Research Cooperative Program, 
  Grant-in-Aid for  Scientific
Research Fund of the Ministry of Education, Science and Culture of Japan 
 No.18540262 and No.17340075.  

\end{acknowledgments}

\end{document}